\documentclass{nlaauth}
\usepackage{graphicx,wrapfig,bm}
\usepackage{pifont,latexsym,ifthen,theorem,rotating,calc}
\usepackage{amsfonts,amssymb,amsbsy,amsmath}
\newtheorem{prop}{Proposition}[section]
\begin{document}
\NLA{1}{11}{xxx}{yyy}{07}
%\NLA{<first page>}{<last page>}{<volume>}{<issue>}{<year (two digit)>}
\runningheads{G.~Ito}{Approximate formulation of the probability}
%\received{<Date>}
%\revised{<Date>}
%\accepted{<Date>}
\noreceived{}
\norevised{}
\noaccepted{}
\title{Approximate formulation of the probability that the Determinant or Permanent of a matrix undergoes the least change under perturbation of a single element}
\author{Genta~Ito\corrauth}
\address{Maruo Lab., 500 El Camino Real \#302, Burlingame, CA 94010, United States.}
\corraddr{Maruo Lab., 500 El Camino Real \#302, Burlingame, CA 94010, United States. (Email: gito@maruolab.com)}
\begin{abstract}
%\chapter{Approximate formulation of the probability that the Determinant or Permanent of a matrix undergoes the least change under perturbation of a single element~\cite{Ito1}}\label{ChapApprox}
In an earlier paper, we discussed the probability that the determinant of a matrix undergoes the least change upon perturbation of one of its elements, provided that most or all of the elements of the matrix are chosen at random and that the randomly chosen elements have a fixed probability of being non-zero. In this paper, we derive approximate formulas for that probability by assuming that the terms in the permanent of a matrix are independent of one another, and we apply that assumption to several classes of matrices. In the course of deriving those formulas, we identified several integer sequences that are not listed on Sloane's Web site.  
\end{abstract}
%%%%%%%%%%%%%%%%%%%%%%%%%%%%%%%%%%%%%%%%%%%%%%%%%%%%%%%%%%%%%%%%%%%
\section{Introduction}\label{SecIntro_p1}
In an earlier paper~\cite{Ito2}, we discussed the problem of finding the probability that the determinant of a matrix undergoes the least change under perturbation of one of its elements. In this paper, we consider only the case where the randomly chosen matrix elements are values of a continuous (real) random variable, and we derive approximate formulas for that probability by assuming that the terms in the permanent of a matrix are mutually independent.
Denote the determinant of any matrix $A$ by $\det A$. For an $(n+1)\times(n+1)$ matrix $M_{n+1}$, the expansion of $\mathrm{det}$\,$M_{n+1}$ via row $i$ is
\begin{equation}\label{DefDeterminant1_p1}
\mathrm{det}\, M_{n+1}=m_{i1}M_{i1}+\cdots+m_{ij}M_{ij}+\cdots+m_{in+1}M_{in+1},
\end{equation}
where $m_{ij}$ is the element of $M_{n+1}$ at the intersection of row $i$ and column $j$, and $M_{ij}$ is the cofactor of $m_{ij}$. 
$M_{ij}$ can be written as $(-1)^{i+j}\det S_{n}$, where $S_{n}$ is the $n\times n$ submatrix of $M_{n+1}$ which is obtained by deleting row $i$ and column $j$. Thus the probability that $\det M_{n+1}$ undergoes the least change upon perturbation of element $m_{ij}$ is equal to the probability that $|\det S_{n}|$ is as small as possible.
As in the earlier paper, we treat the following three classes of $n\times n$ matrices $S_{n}$:
\begin{enumerate}
\item[(i)]Matrices $A_{n}$ in which all the elements are values of mutually independent random variables each of which has probability $r$ of being non-zero and probability $1-r$ of being 0, where $0<r<1$.
\item[(ii)]Matrices $B_{n}$ in which all but one of the diagonal elements are set to 1 (i.e., $b_{ii}=1$ for $i\ne 1$, where $b_{11}$ is the special diagonal element), and $b_{11}$ and all the off-diagonal elements are as in (i). Thus  
\begin{equation}\label{EqMatB_p1}
B_{n}=\left(\begin{array}{ccccc}b_{11}&b_{12}&b_{13}&\cdots& b_{1n}\\
b_{21}&1&b_{23}&\cdots& b_{2n}\\
b_{31}&b_{32}&1&\cdots& b_{3n}\\
\vdots&\vdots&\vdots&\ddots&\vdots\\
b_{n1}&b_{n2}&b_{n3}&\cdots& 1\\\end{array}\right)
\end{equation}
\item[(iii)]Matrices $C_{n}$ in which all the diagonal elements are set to 1 and all the off-diagonal elements are as in (i).  Thus
\begin{equation}\nonumber 
C_{n}=\left(\begin{array}{ccccc}1&c_{12}&c_{13}&\cdots&c_{1n}\\
c_{21}&1&c_{23}&\cdots&c_{2n}\\c_{31}&c_{32}&1&\cdots&c_{3n}\\
\vdots&\vdots&\vdots&\ddots&\vdots\\
c_{n1}&c_{n2}&c_{n3}&\cdots& 1\\\end{array}\right)
\end{equation}
\end{enumerate}
The matrices $S_{n}$ described in (i), (ii), and (iii) above will be said to be matrices of type $\mathbf{A}_n,\mathbf{B}_n$, and $\mathbf{C}_n$, respectively. Every randomly chosen element of a matrix of any of these types (i.e., every element of a matrix of type $\mathbf{A}_n$, every off-diagonal element of a matrix of type $\mathbf{B}_n$ or $\mathbf{C}_n$, and the special diagonal element $b_{11}$ of a matrix of type $\mathbf{B}_n$) will be called a {\it variable} element; the remaining elements will be called {\it fixed} elements.
If $\mathfrak{X}$ is a continuous set such that $0\in\mathfrak{X}$ (as is assumed here), and $X$ is a random variable whose values are selected from $\mathfrak{X}$, then the probability function $P$ for $\mathfrak{X}$ is defined by
\begin{equation}\label{DefProbability_p1}
\begin{array}{rcl}
P(X=0)&=&1-r\\P(X\ne 0)&=&r\\
\end{array}
\end{equation}
Since $\mathfrak{X}$ is continuous, $P$ can be expressed in terms of a density function $q$ as
\begin{equation}\label{ProbContinuousX_p1}
\mbox{$D\subseteq X\Rightarrow P(X\displaystyle \in D)=\int_{D}\,q(x)dx$,}
\end{equation}
where
\begin{equation}\nonumber
0\leqq q(x)\leqq 1,\hspace*{.2in}\mbox{$\displaystyle\int_{\{0\}}q(x)dx=1-r$},\hspace*{.2in}\mbox{$\displaystyle \int_{\mathfrak{X}-\{0\}}\,q(x)dx=r$}
\end{equation}
If $q$ is continuous on $\mathfrak{X}-\{0\}$ (as is assumed here), then 
\begin{equation}\label{FeatureProbOnContinuousX_p1}
P(X=x)=\left\{\begin{array}{rcl}1-r,&\hspace*{.2in}&x=0\\
0,&\hspace*{.2in}&x\in\mathfrak{X}-\{0\}\\
\end{array}\right.
\end{equation}
As found in~\cite{Ito2}, the probability that $\det M_{n+1}$ undergoes the least change upon perturbation of element $m_{ij}$ is equal to the probability that $\det S_{n}=u$, where $u=0$ if $S_{n}$ is of type $\mathbf{A}_n$ or $\mathbf{B}_n$, and $u=1$ if $S_{n}$ is of type $\mathbf{C}_n$.
In~\cite{Ito2}, the probability that $\det S_{n}=u$ was formulated in terms of binary matrices (matrices of 0's and 1's), as follows:
For a matrix $S_{n}$ of type $\mathbf{A}_n,\ \mathbf{B}_n$, or $\mathbf{C}_n,\ \tilde{S}_{n}$ is the $n\times n$ binary matrix with $\tilde{s}_{ij}=1$ if $s_{ij}\ne 0$, and $\tilde{s}_{ij}=0$ if $s_{ij}=0$. Moreover, $\tilde{s}_{ij}$ is a {\it variable} (resp.\ {\it fixed}) element of $\tilde{S}_{n}$ if $s_{ij}$ is a variable (resp.\ fixed) element of $S_{n}$, and $\tilde{S}_{n}$ is of type $\tilde{\mathbf{A}}_{n}$ (resp.\ $\tilde{\mathbf{B}}_{n},\ \tilde{\mathbf{C}}_{n}$) if $S_{n}$ is of type $\mathbf{A}_n$ (resp.\ $\mathbf{B}_n,\ \mathbf{C}_n$).  Every variable element of $\tilde{S}_{n}$ has probability $r$ of being 1, and probability $1-r$ of being 0.
Denote the permanent of any matrix $A$ by $\mathrm{per}\, A$. The expansion of the permanent of $\tilde{S}_{n}$ via the permutation group $\mathfrak{S}_n$ on the set $\{1,2,3,\ldots,n\}$ is
\begin{equation}\label{DefPermanent_p1}
\displaystyle \mathrm{per}\,\tilde{S}_{n}=\sum_{\sigma\in\mathfrak{S}_{n}}\tilde{s}_{\sigma(1)1}\tilde{s}_{\sigma(2)2}\cdots\tilde{s}_{\sigma(n)n}
\end{equation}
Since $\tilde{S}_{n}$ is a binary matrix, every term in this expansion is either 0 or 1, so the value of $\mathrm{per}\,\tilde{S}_{n}$ is a non-negative integer.
In~\cite{Ito2}, it was shown that the probability that $\det S_{n}=u$ is equal to the probability that $\mathrm{per}\,\tilde{S}_{n}=u$, and that $\tilde{S}_{n}$ has non-zero (positive) probability of having permanent $u$ if and only if every term in the expansion of $\mathrm{per}\,\tilde{S}_{n}$ via $\mathfrak{S}_n$ that contains at least one variable element is 0. Thus if $S_{n}$ is of type $\mathbf{A}_n$ or $\mathbf{B}_n$, then $S_{n}$ has non-zero probability of having determinant 0 if and only if {\it every} term in the expansion of $\mathrm{per}\,\tilde{S}_{n}$ is 0; and if $S_{n}$ is of type $\mathbf{C}_n$, then $S_{n}$ has non-zero probability of having determinant 1 if and only if the only non-zero term in the expansion of $\mathrm{per}\,\tilde{S}_{n}$ is the ``diagonal'' term, $\tilde{c}_{11}\tilde{c}_{22}\tilde{c}_{33}\cdots\tilde{c}_{nn}$.
A matrix of type $\tilde{\mathbf{A}}_{n}$ can be converted to a matrix of type $\tilde{\textbf{B}}_{n}$ by replacing all but one of the diagonal elements with a fixed element 1.  The latter matrix can in turn be converted to a matrix of type $\tilde{\mathbf{C}}_{n}$ by replacing the sole variable element on the main diagonal (element $\tilde{b}_{11}$) with a fixed element 1.  Therefore, we might expect to find that there exist analytic formulas for $P_{\mathrm{per}\,\tilde{B}_{n}=0}(r)$ in terms of $P_{\mathrm{per}\,\tilde{A}_{n}=0}(r)$, and $P_{\mathrm{per}\,\tilde{C}_{n}=1}(r)$ in terms of $P_{\mathrm{per}\,\tilde{B}_{n}=0}(r)$, where $P_{\mathrm{per}\,\tilde{S}_{n}=u}(r)$ denotes the probability that $\mathrm{per}\,\tilde{S}_{n}=u$.
In~\cite{Ito2}, the exact probability $P_{\mathrm{per}\,\tilde{S}_{n}=u}(r)$ for a specific type of matrix ($\tilde{\mathbf{A}}_{n},\ \tilde{\mathbf{B}}_{n}$, or $\tilde{\mathbf{C}}_{n}$) was formulated in terms of the numbers of matrices $\tilde{S}_{n}$ of that type which have $i$ variable elements with a value of 1 in the expansion of their permanent (where $i$ ranged from 0 to some $i_{\max}$ that depended on the type of matrix). In this paper, we will derive an approximate probability $Q_{\mathrm{per}\,\tilde{S}_{n}=u}(r)$ that $\mathrm{per}\,\tilde{S}_{n}=u$, by assuming that all the terms in the expansion of $\mathrm{per}\,\tilde{S}_{n}$ via $\mathfrak{S}_n$ are independent of one another.  For example, the expansion of the permanent of a matrix of type $\tilde{\mathbf{A}}_{3}$ is
\begin{eqnarray*}
\mathrm{per}\tilde{A}_{3}&=&\tilde{a}_{11}\tilde{a}_{22}\tilde{c}_{33}+\tilde{a}_{13}\tilde{a}_{32}\tilde{a}_{21}+\tilde{a}_{12}\tilde{a}_{23}\tilde{a}_{31}+\tilde{a}_{13}\tilde{a}_{31}\tilde{a}_{22}\\&&\qquad+\tilde{a}_{11}\tilde{a}_{23}\tilde{a}_{32}+\tilde{a}_{12}\tilde{a}_{21}\tilde{a}_{33}
\end{eqnarray*}
If $\tilde{a}_{23}=0$, the terms $\tilde{a}_{12}\tilde{a}_{23}\tilde{a}_{31}$ and $\tilde{a}_{11}\tilde{a}_{23}\tilde{a}_{32}$ must both be zero, but we will ignore that relationship. Instead, for every $m\le n$ and each of the three types of matrices ($\tilde{\mathbf{A}}_{n},\ \tilde{\mathbf{B}}_{n}$, and $\tilde{\mathbf{C}}_{n}$), we will consider only the number of terms in the expansion of the permanent of a matrix of that type which have $m$ variable elements, regardless of the values of those variable elements or the connections between different terms in the expansion.
\section{Formulation of $Q_{\mathrm{per}\,\tilde{S}_{n}=u}(r)$}\label{SecCases_p1}
There are $n!$ terms $\tilde{s}_{\sigma(1)1}\tilde{s}_{\sigma(2)2}\cdots\tilde{s}_{\sigma(n)n}$ in (\ref{DefPermanent_p1}), because the number of elements $\sigma\in\mathfrak{S}_{n}$ is $n!$.  For each term, let $m$ be the number of variable elements it contains, and let $E_{n}(m)$ be the number of terms with $m$ variable elements.  Then
\begin{equation}\nonumber
Q_{\mathrm{per}\,\tilde{S}_{n}=u}(r)=\displaystyle \prod_{m=1}^{n}(1-r^{m})^{E_{n}(m)},
\end{equation}
where $r^{m}$ is the probability that all the variable elements in a term with $m$ variable elements are non-zero, $1-r^{m}$ is the probability that at least one variable element in a term with $m$ variable elements is 0, and $(1-r^{m})^{E_{n}(m)}$ is the probability that all the terms with $m$ variable elements are 0.
What remains to be done is to determine $E_{n}(m)$ for every $m\le n$ and each of the three types of matrices.
\subsection{Type $\tilde{\mathbf{A}}_{n}$}\label{SubSecCase(i)_p1}
For type $\tilde{\mathbf{A}}_{n}$, every term in the expansion of $\mathrm{per}\,\tilde{A}_{n}$ has $n$ variable elements, so $E_{n}(m)=0$ for every $m<n$. Since there are $n!$ terms, we obtain
\begin{equation}\label{Acurve_p1}
Q_{\mathrm{per}\,\tilde{A}_{n}=0}(r)=(1-r^{n})^{n!}
\end{equation}
\subsection{Type $\tilde{\mathbf{C}}_{n}$}\label{SubSecCase(iii)_p1}
For type $\tilde{\mathbf{C}}_{n}$, let $W_{n}(m)$ denote $E_{n}(m)$, so that
\begin{equation}\label{EqQC_p1}
Q_{\mathrm{per}\,\tilde{C}_{n}=1}(r)=\displaystyle \prod_{m=1}^{n}(1-r^{m})^{W_{n}(m)}
\end{equation}
\begin{prop}
\label{PropHn1_p1}
\begin{equation}\label{EqWn1_p1}
W_{n}(n-1)=n\cdot W_{n-1}(n-1)
\end{equation}
\end{prop}
{\bf Proof}\ \ \ 
Every term in the expansion of $\mathrm{per}\,\tilde{C}_{n}$ that has $n-1$ variable elements contains just one fixed (diagonal) element (namely, $\tilde{c}_{ii}$ for some $i$ with $1\le i\le n$).
\begin{figure}[h]
\begin{center}
\includegraphics[width=90mm,height=24.4mm]{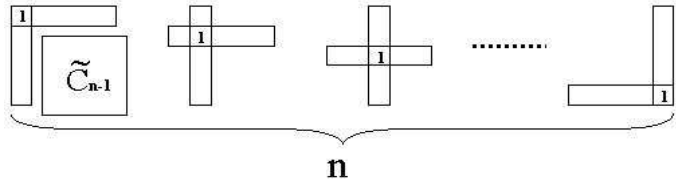}
\end{center}
\caption{Cross shapes deleted from $\tilde{C}_n$ to create the $(n-1)\times(n-1)$ submatrices $\tilde{C}^i_n$ of $\tilde{C}_{n}$}
\label{FigWn1_p1}\end{figure}
For every $i$, there exists a natural one-to-one correspondence between the set $T_{1i}$ of terms in the expansion of $\mathrm{per}\,\tilde{C}_{n}$ that contain element $\tilde{c}_{ii}$ (and have $n-1$ variable elements of $\tilde{C}_{n}$ apiece) and the set $T_{2i}$ of terms in the expansion of $\mathrm{per}\,\tilde{C}_{n}^{i}$ that have $n-1$ variable elements each, where $\tilde{C}_{n}^{i}$ is the $(n-1)\times(n-1)$ submatrix of $\tilde{C}_{n}$ which is formed by deleting the ``cross shape'' that consists of row $i$ and column $i$. There are $n$ such cross shapes (one for each diagonal element of $\tilde{C}_{n}$), as shown in Fig.~\ref{FigWn1_p1}; moreover, for every $i,\ \\tilde{C}_{n}^{i}$ is a matrix of type $\tilde{\mathbf{C}}_{n-1}$. Thus we obtain (\ref{EqWn1_p1}).  $\blacksquare$
%%%%%%%%%%%%%%%%%%%%%%%%%%%%%%%%%%%%%%%%%%%%%%%%%%%%%%%%%%%%%%%%%%%%
\begin{prop}\label{PropWn2_p1}
For every $m$ with $1\leqq m\leqq n-1$,
\begin{equation}\label{EqWn2_1_p1}
W_{n}(m)=\displaystyle \frac{{}_n\mathrm{P}_m}{m!}\cdot W_{m}(m)
\end{equation}
\end{prop}
{\bf Proof}\ \ \ 
Let $m,\,k$ be such that $1\le m\le n-1$ and $k=n-m$. A term in the expansion of $\mathrm{per}\,\tilde{C}_{n}$ has $m$ variable elements if and only if it has $k$ fixed (diagonal) elements (namely, $\tilde{c}_{i_{1}i_{1}},\ldots,\tilde{c}_{i_{k}i_{k}}$ for some $i_{1},\ldots,i_{k}$ with $1\le i_{1}<i_{2}<\cdots <i_{k}\le n$). For every $k$-tuple $(i_{1},\ldots,i_{k})$ with $1\le i_{1}<i_{2}<\cdots<i_{k}\le n$, there exists a natural one-to-one correspondence between the set $T_{1,\{i_{1},\ldots,i_{k}\}}$ of terms in the expansion of $\mathrm{per}\,\tilde{C}_{n}$ that contain elements $\tilde{c}_{i_{1}i_{1}},\ldots,\tilde{c}_{i_{k}i_{k}}$ (and have $m$ variable elements of $\tilde{C}_{n}$ apiece) and the set $T_{2,\{i_{1},\ldots,i_{k}\}}$ of terms in the expansion of $\mathrm{per}\,\tilde{C}_{n}^{\{i_{1},\ldots,i_{k}\}}$ that have $m$ variable elements each, where $\tilde{C}_{n}^{\{i_{1},\ldots,i_{k}\}}$ is the $m\times m$ submatrix of $\tilde{C}_{n}$ which is formed by deleting the cross shapes that correspond to $i_{1},\ldots,i_{k}$ (i.e., the submatrix of $\tilde{C}_{n}$ which is formed by deleting rows $i_{1},\ldots,i_{k}$ and columns $i_{1},\ldots,i_{k}$). The number of $k$-tuples $(i_{1},\ldots,i_{k})$ with $1\le i_{1}<i_{2}<\cdots <i_{k}\le n$ is ${}_n\mathrm{C}_k$; moreover, for each such $k$-tuple, $\\tilde{C}_{n}^{\{i_{1},\ldots,i_{k}\}}$ is a matrix of type $\tilde{\mathbf{C}}_{m}$. Also, ${}_n\mathrm{C}_k=\displaystyle \frac{{}_n\mathrm{P}_m}{m!}$, since $k=n-m$ and ${}_n\mathrm{C}_{n-m}=\displaystyle \frac{n!}{(n-m)!m!}=\frac{{}_n\mathrm{P}_m}{m!}$. Thus we obtain (\ref{EqWn2_1_p1}). $\blacksquare$
%%%%%%%%%%%%%%%%%%%%%%%%%%%%%%%%%%%%%%%%%%%%%%%%%%%%%%%%%%%%%%%%%%%%
\begin{prop}\label{PropWn3_p1}
For $n\ge 1$,
\begin{equation}
\label{EqWn3_1_p1}
W_{n}(n)=\displaystyle \sum_{j=1}^{n-1}(-1)^{n-j+1}\cdot{}_{n}\mathrm{P}_{j-1}
\end{equation}
\end{prop}
{\bf Proof}\ \ \ $W_{n}(n)$ is the number of permutations of the numbers $1,2,3,\ldots,n$ with no fixed points (i.e., the number of permutations that have no cycles of length one).  Such permutations are called {\it derangements}. In 1713, Pierre de Montmort~\cite{Montmort}   proved that
\begin{equation}\label{EqHn3_2_p1}
W_{n}(n)=n\cdot W_{n-1}(n-1)+(-1)^{n}
\end{equation}
Thus we obtain (\ref{EqWn3_1_p1}) recursively as follows:
\begin{eqnarray}\label{EqWn3_2_p1}
 W_{n}(n)&=&n\cdot W_{n-1}(n-1)+(-1)^{n}\\\nonumber
&=&{}_{n}\mathrm{P}_{1}\cdot W_{n-1}(n-1)+(-1)^{n}\cdot{}_{n}\mathrm{P}_{0}\\\nonumber
&=&n\cdot\left[(n-1)W_{n-2}(n-2)+(-1)^{n-1}\right]+(-1)^{n}\\\nonumber
&=&n(n-1)\cdot W_{n-2}(n-2)+n\cdot(-1)^{n-1}+(-1)^{n}\\\nonumber
&=&{}_{n}\mathrm{P}_{2}\displaystyle \cdot W_{n-2}(n-2)+\sum_{j=1}^{2}(-1)^{n-j+1}\cdot{}_{n}\mathrm{P}_{j-1}\\\nonumber
&=&\displaystyle \cdots={}_{n}\mathrm{P}_{k}\cdot W_{n-k}(n-k)+\sum_{j=1}^{k}(-1)^{n-j+1}\cdot{}_{n}\mathrm{P}_{j-1}\\\nonumber
&=&\displaystyle \cdots={}_{n}\mathrm{P}_{n-1}\cdot W_{1}(1)+\sum_{j=1}^{n-1}(-1)^{n-j+1}\cdot{}_{n}\mathrm{P}_{j-1}\\\nonumber
&=&\displaystyle \sum_{j=1}^{n-1}(-1)^{n-j+1}\cdot{}_{n}\mathrm{P}_{j-1}
\end{eqnarray}
The last step follows from the fact that $W_{1}(1)=0$.  $\blacksquare$
%%%%%%%%%%%%%%%%%%%%%%%%%%%%%%%%%%%%%%%%%%%%%%%%%%%%%%%%%%%%%%%%%%%%
\begin{prop}\label{PropWn4_p1}
For every $m$ with $0\leq m\leq n$,
\begin{equation}
\label{EqWn4_1_p1}
W_{n}(m)={}_{n}\mathrm{P}_{m}\displaystyle \cdot\sum_{l=0}^{m}\frac{(-1)^{l}}{l!}
\end{equation}
\end{prop}
{\bf Proof}\ \ \ The only term in the expansion of $\mathrm{per}\,\tilde{C}_{n}$ that has no variable elements is $\tilde{c}_{11}\cdots\tilde{c}_{nn}$, so $W_{n}(0)=1={}_{n}\mathrm{P}_{0}\displaystyle \cdot\frac{(-1)^{0}}{0!}$. This proves that~(\ref{EqWn4_1_p1}) holds for $m=0$. By~(\ref{EqWn3_1_p1}),
\begin{eqnarray}
\displaystyle \nonumber W_{n}(n)&=&\displaystyle \sum_{j=1}^{n-1}(-1)^{n-j+1}\cdot{}_{n}\mathrm{P}_{j-1}\\\nonumber&=&\displaystyle \sum_{j=1}^{n-1}(-1)^{n-j+1}\cdot\frac{n!}{(n-j+1)!}\\\nonumber&=&n!\displaystyle \cdot\sum_{j=1}^{n-1}\frac{(-1)^{n-j+1}}{(n-j+1)!}={}_n\mathrm{P}_{n}\cdot\sum_{l=2}^{n}\frac{(-1)^{l}}{l!}\\\nonumber&=&{}_n\mathrm{P}_{n}\displaystyle \cdot\sum_{l=0}^{n}\frac{(-1)^{l}}{l!}
\end{eqnarray}
 This proves that~(\ref{EqWn4_1_p1}) holds for $m=n$ (the last step follows from the fact that $\displaystyle \sum_{l=0}^{1}\frac{(-1)^{l}}{l!}=1-1=0$), so we can assume that $1\le m\le n-1$. By (\ref{EqWn2_1_p1}) and (\ref{EqWn3_1_p1}),
\begin{eqnarray}\nonumber
 W_{n}(m)&=&\displaystyle \frac{{}_{n}\mathrm{P}_{m}}{m!}\cdot W_{m}(m)=\frac{\left[\frac{n!}{(n-m)!}\right]}{m!}\cdot\sum_{j=1}^{m-1}(-1)^{m-j+1}\cdot{}_{m}\mathrm{P}_{j-1}\\\nonumber
&=&\displaystyle \frac{\left[\frac{n!}{(n-m)!}\right]}{m!}\cdot\sum_{j=1}^{m-1}(-1)^{m-j+1}\cdot\frac{m!}{(m-j+1)!}\\\nonumber
&=&\displaystyle \frac{n!}{(n-m)!}\cdot\sum_{j=1}^{m-1}\frac{(-1)^{m-j+1}}{(m-j+1)!}={}_{n}\mathrm{P}_{m}\cdot\sum_{l=2}^{m}\frac{(-1)^{l}}{l!}\\\nonumber
&=&{}_{n}\mathrm{P}_{m}\displaystyle \cdot\sum_{l=0}^{m}\frac{(-1)^{l}}{l!}
\end{eqnarray}$\blacksquare$
The values of $W_{n}(m)$ for $n=1,\ldots,6$ (and $m=0,\ldots,n$) are given in Table~\ref{TableWn_p1}.
\begin{table}[ht]
    \scriptsize
\begin{center}
    \caption{Values of $W_n(m)$ for $n=1,\ldots,6$ (and $m=0,\ldots,n$)}
\begin{tabular}{|c|ccccccccccccc|}\hline
     $n\backslash m$&0&&1&&2&&3&&4&&5&&6\\ \hline
    1&1&$\underrightarrow{${\scriptsize $-1$}$}$&0&&&&&&&&&&\\
    &$\downarrow^{\times\frac{2}{2}}$&&$\downarrow^{\times 2}$&&&&&&&&&&\\
    2&1&&0&$\underrightarrow{${\scriptsize $+1$}$}$&1&&&&&&&&\\
    &$\downarrow^{\times\frac{3}{3}}$&&$\downarrow^{\times\frac{3}{2}}$&&$\downarrow^{\times 3}$&&&&&&&&\\
    3&1&&0&&3&$\underrightarrow{${\scriptsize $-1$}$}$&2&&&&&&\\
    &$\downarrow^{\times\frac{4}{4}}$&&$\downarrow^{\times\frac{4}{3}}$&&$\downarrow^{\times\frac{4}{2}}$&&$\downarrow^{\times 4}$&&&&&&\\
    4&1&&0&&6&&8&$\underrightarrow{${\scriptsize $+1$}$}$&9&&&&\\
    &$\downarrow^{\times\frac{5}{5}}$&&$\downarrow^{\times\frac{5}{4}}$&&$\downarrow^{\times\frac{5}{3}}$&&$\downarrow^{\times\frac{5}{2}}$&&$\downarrow^{\times 5}$&&&&\\
    5&1&&0&&10&&20&&45&$\underrightarrow{${\scriptsize $-1$}$}$&44&&\\
    &$\downarrow^{\times\frac{6}{5}}$&&$\downarrow^{\times\frac{6}{5}}$&&$\downarrow^{\times\frac{6}{4}}$&&$\downarrow^{\times\frac{6}{3}}$&&$\downarrow^{\times\frac{6}{2}}$&&$\downarrow^{\times 6}$&&\\
    6&1&&0&&15&&40&&135&&264&$\underrightarrow{${\scriptsize $+1$}$}$&265\\
    &$\downarrow^{\times\frac{7}{7}}$&&$\downarrow^{\times\frac{7}{6}}$&&$\downarrow^{\times\frac{7}{5}}$&&$\downarrow^{\times\frac{7}{4}}$&&$\downarrow^{\times\frac{7}{3}}$&&$\downarrow^{\times\frac{7}{2}}$&&$\downarrow^{\times 7}$\\ \hline
\end{tabular}
\label{TableWn_p1}
\end{center}
\end{table}
The sequence $\{W_{n}(n)\}_{n\ge 0}$ is number A000166 in Sloane's list~\cite{Sloane}.  The first ($n=0$) term, which has no meaning for our purposes, is given there as 1. For $n\ge 1$,  $W_{n}(n)$ is not only the number of derangements of the numbers $1,2,3,\ldots,n$, but also the permanent of the $n\times n$ binary matrix with 0's on the main diagonal and 1's everywhere else.  The sequences $\{W_{n}(2)\}_{n\ge 2},\,\{W_{n}(3)\}_{n\ge 3},\,\{W_{n}(4)\}_{n\ge 4}$, and $\{W_{n}(5)\}_{n\ge 5}$ are given in Sloane's list (numbers A00217, A007290, A060008, and A060836, respectively).  There is no $m>5$ for which the sequence $\{W_{n}(m)\}_{n\ge m}$ is given in that list.   
An alternative method of determining the value of $W_{n}(m)$ is to use the cycle structure of the permutation group $\mathfrak{S}_n$.  For $n=5$, the types and numbers of cycles associated with elements of $\mathfrak{S}_5$ that are applicable to each value of $m$ are listed in Table~\ref{TableS5_p1}, together with the values of $W_{5}(m)$ computed from them.  For $m=4$ and $m=5$, there are two different types of cycles each. The permutations that correspond to $m=4$ (which index the terms in the expansion of $\mathrm{per}\,\tilde{S}_{n}$ that have 4 variable elements) are those that can be expressed as a 4-cycle (such as $(1234)$) and those that can be expressed as a product of two 2-cycles (such as $(12)(34)$).  Similarly, the permutations that correspond to $m=5$ (which index the terms with 5 variable elements) are the one that can be expressed as the 5-cycle $(12345)$ and those that can be expressed as a product of one 2-cycle and one 3-cycle (such as $(12)(345)$).
\begin{table}[ht]
    \scriptsize
\begin{center}
    \caption{Computation of $W_5(m)$ from group table for $\mathfrak{S}_5$}
\begin{tabular}{|c|c|c|c|c|}\hline
$m$& cycle & rep. & No.\ of cycles & $W_{5}(m)$  \\ \hline
    $0$&$1^{5}$&$e$&$5!/\left(1^{5}\cdot 5!\right)=1$&1\\ \hline
    $2$&$2^{1}\cdot 1^{3}$&$(12)$&$5!/\left(2^{1}\cdot 1!\cdot 1^{3}\cdot 3!\right)=10$&$10$\\ \hline
    $3$&$3^{1}\cdot 1^{2}$&$(123)$&$5!/\left(3^{1}\cdot 1!\cdot 1^{2}\cdot 2!\right)=20$&$20$\\ \hline
    $4$&$4^{1}\cdot 1^{1}$&$(1234)$&$5!/\left(4^{1}\cdot 1!\cdot 1^{1}\cdot 1!\right)=30$&$30+15$\\ \cline{2-4}
    &$2^{2}\cdot 1^{1}$&$(12)(34)$&$5!/\left(2^{2}\cdot 2!\cdot 1!\cdot 1^{1}\right)=15$&$=45$\\ \hline
    $5$&$5^{1}$&$(12345)$&$5!/\left(5^{1}\cdot 1!\right)=24$&$24+20$\\ \cline{2-4}
    &$2^{1}\cdot 3^{1}$&$(12)(345)$&$5!/\left(2^{1}\cdot 1!\cdot 3^{1}\cdot 1!\right)=20$&$=44$\\ \hline
\end{tabular}
\label{TableS5_p1}   
\end{center}
\end{table}
\subsection{Type $\tilde{\mathbf{B}}_{n}$}\label{SubSecCase(ii)_p1}
For type $\tilde{\mathbf{B}}_{n}$, let $V_{n}(m)$ denote $E_{n}(m)$, so that
\begin{equation}\label{EqQB_p1}
Q_{\mathrm{per}\,\tilde{B}_{n}=0}=\displaystyle \prod_{m=1}^{n}(1-r^{m})^{V_{n}(m)}
\end{equation}
\begin{prop}\label{PropVn}   For every $m$ with $1\leq m\leq n$,
\begin{equation}
\label{EqGn2_p1}
V_{n}(m)=\displaystyle \frac{{}_{n}\mathrm{P}_{m}}{n}\cdot\left\{\left[(m+1)\cdot\sum_{l=0}^{m}\frac{(-1)^{l}}{l!}\right]-\frac{(-1)^{m}}{m!}\right\}
\end{equation}
\end{prop}
{\bf Proof}\ \ \ 
Let $\tilde{b}_{11}$ be the sole variable element on the main diagonal of a matrix of type $\tilde{\mathbf{B}}_{n}$. If we delete the L-shape that consists of row $1$ and column $1$, we obtain a matrix of type $\tilde{\mathbf{C}}_{n-1}$.  Therefore, 
\begin{equation}\label{EqVnWnRecurrent_p1}
V_{n}(m)-W_{n-1}(m-1)=W_{n}(m)-W_{n-1}(m)
\end{equation}
By (\ref{EqWn4_1_p1}),
\begin{eqnarray}\nonumber
V_{n}(m) &=& W_{n}(m)-W_{n-1}(m)+W_{n-1}(m-1) \\ \nonumber
&=&\left[ {}_{n}\mathrm{P}_{m} \cdot \displaystyle \sum_{l=0}^{m}\frac{(-1)^{l}}{l!}\right] - \left[ {}_{n-1}\mathrm{P}_{m} \cdot \displaystyle \sum_{l=0}^{m}\frac{(-1)^{l}}{l!}\right] +\left[ {}_{n-1}\mathrm{P}_{m-1} \cdot \displaystyle \sum_{l=0}^{m-1}\frac{(-1)^{l}}{l!}\right] \\ \nonumber
&=& \displaystyle \frac{{}_{n}\mathrm{P}_{m}}{n}\cdot\left\{\left[(m+1)\displaystyle \cdot\sum_{l=0}^{m}\frac{(-1)^{l}}{l!}\right]-\displaystyle \frac{(-1)^{m}}{m!}\right\} \ \ \ \blacksquare
\end{eqnarray}
The values of $V_{n}(m)$ for $n=1,\ldots,8$ (and $m=0,\ldots,n$) are given in Table~\ref{TableVn_p1}.
\begin{table}[htbp]
    \scriptsize
\begin{center}
\caption{Values of $V_{n}(m)$ for $ n=1,\ldots,8$ (and $m=0,\ldots,n$)}
\begin{tabular}{|c|c|c|c|c|c|c|c|c|c|}\hline
$n\backslash m$&0&1&2&3&4&5&6&7&8\\ \hline
    1&0&1&&&&&&&\\ \hline
    2&0&1&1&&&&&&\\ \hline
    3&0&1&2&3&&&&&\\ \hline
    4&0&1&3&9&11&&&&\\ \hline
    5&0&1&4&18&44&53&&&\\ \hline
    6&0&1&5&30&110&265&309&&\\ \hline
    7&0&1&6&45&220&795&1854&2119&\\ \hline
8&0&1&7&63&385&1855&6489&14833&16687\\ \hline
\end{tabular}
\label{TableVn_p1}
\end{center}
\end{table}
The sequence $\{V_{n}\}_{n\ge 0}$ is number A000255 in Sloane's list~\cite{Sloane}.  The first ($n=0$) term, which has no meaning for our purposes, is given there as 1. For $n\ge 1,\,\,V_{n}(n)$ is the number of permutations of the numbers $1,2,3,\ldots,n+1$ such that there is no element $k$ which is mapped to $k+1$; it is also the permanent of the $n\times n$ binary matrix that has 0's in all but one of the elements on the main diagonal, and 1's everywhere else.  The only value of $m$ for which the sequence $\{V_{n}(m)\}_{n\ge m}$ is given in Sloane's list is 3 (where the sequence $\{V_{n}(3)\}_{n\ge 3}$ is number A045943).  
\section{Comparison of Formulas}\label{SecRemarks_p1}
\begin{figure}[h]
\begin{center}
\includegraphics[width=76.8mm,height=44.2mm]{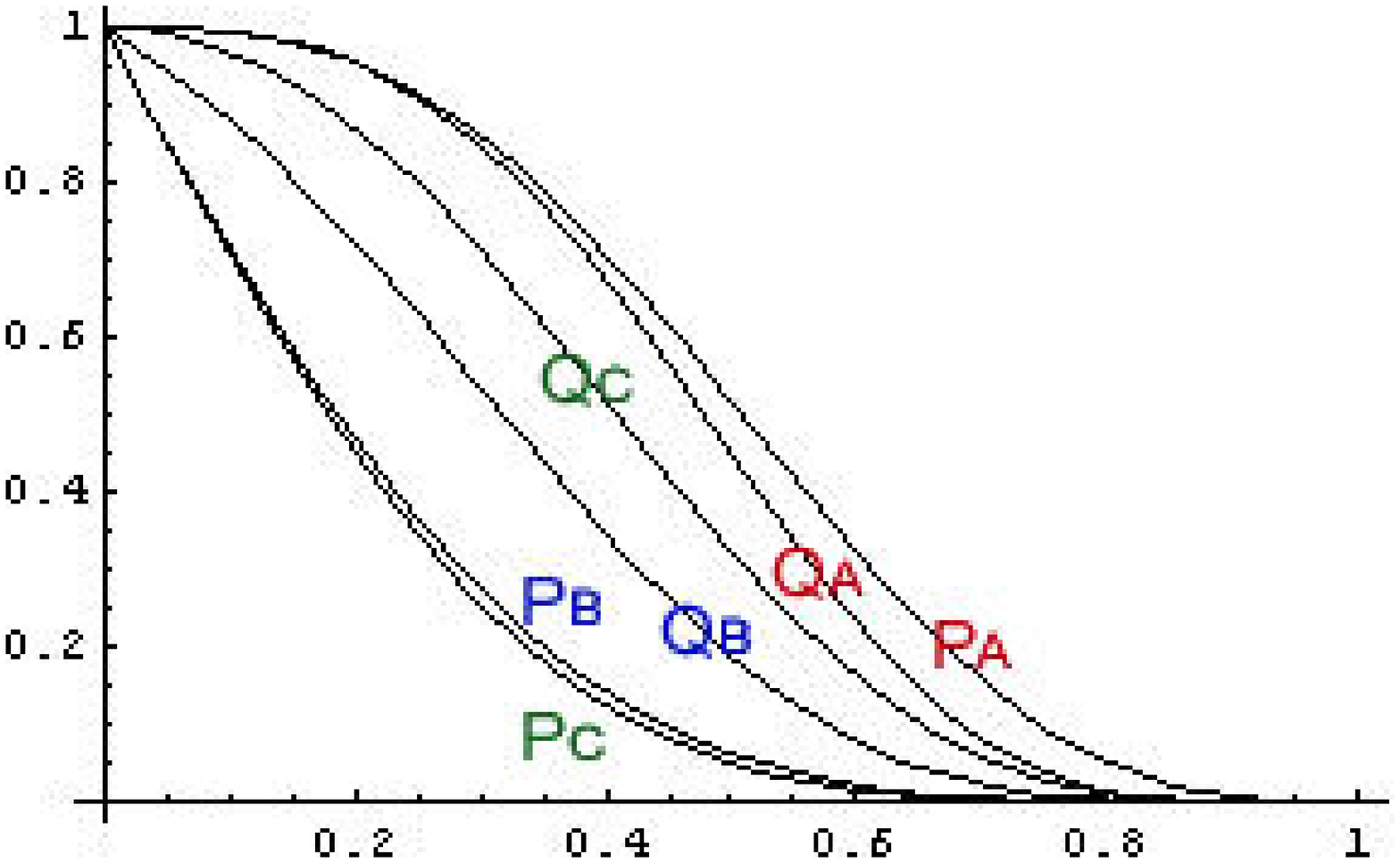}
\end{center}
\caption{Graphs of $Q_{\mathrm{per}\,\tilde{S}_3=u}(r)$ and $P_{\mathrm{per}\,\tilde{S}_3=u}(r)$ vs.\ $r$}
\label{Fig6curves_n3_p1}\end{figure}

\begin{figure}[h]
\begin{center}
\includegraphics[width=76.8mm,height=44.2mm]{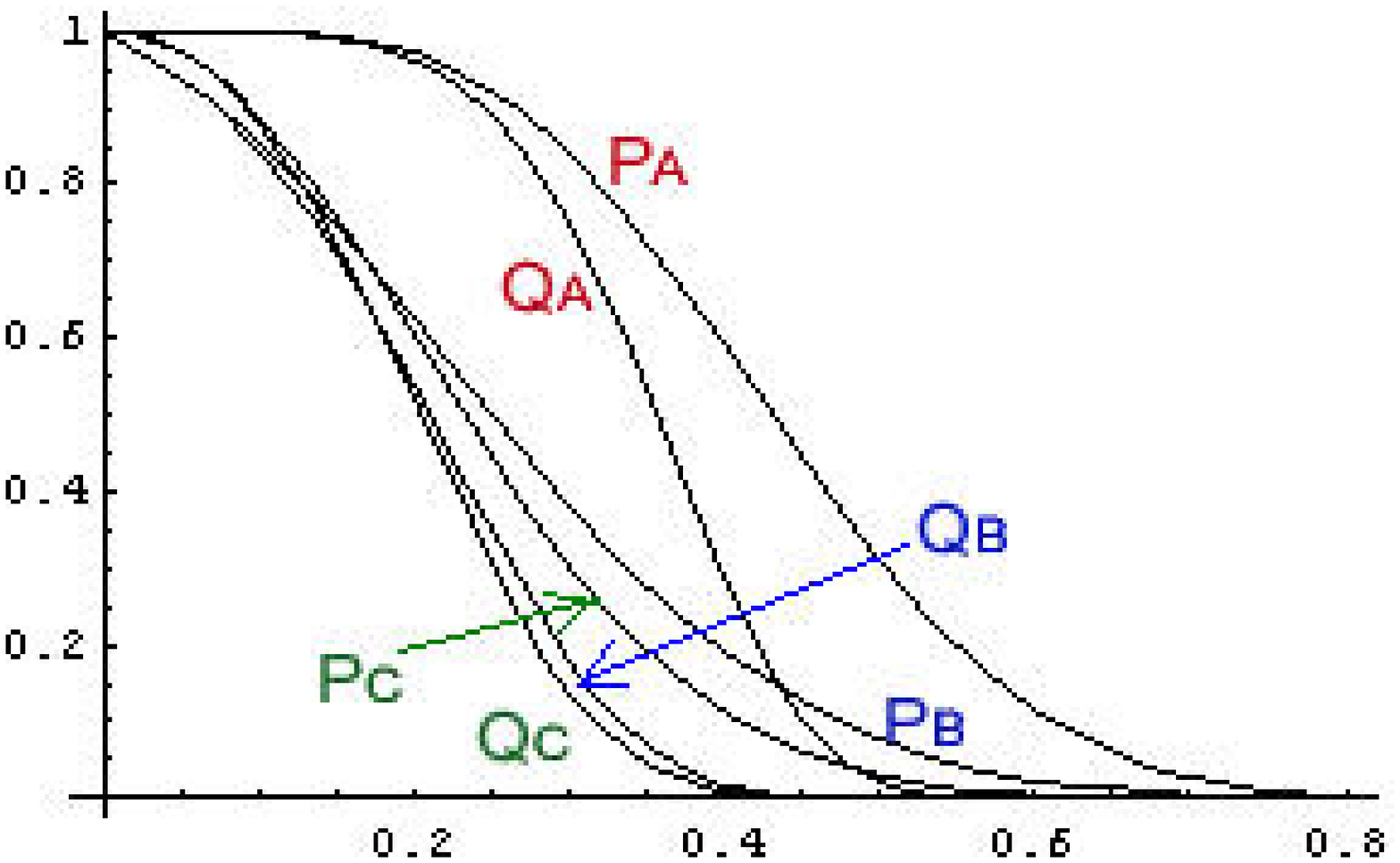}
\end{center}
\caption{Graphs of $Q_{\mathrm{per}\,\tilde{S}_5=u}(r)$ and $P_{\mathrm{per}\,\tilde{S}_5=u}(r)$ vs.\ $r$}
\label{Fig6curves_n5_p1}\end{figure}

For $n=3$, the formulas for the approximate probabilities $Q_{\mathrm{per}\,\tilde{S}_{3}=u}(r)$ are
\begin{eqnarray}\nonumber
 Q_{\mathrm{per}\,\tilde{A}_{3}=0}(r)&=&(1-r{}^{3})^{3!}=(1-r^{3})^{6}\\\nonumber
 Q_{\mathrm{per}\,\tilde{B}_{3}=0}(r)&=&\displaystyle \prod_{m=1}^{3}(1-r^{m})^{V_{3}(m)}=(1-r^{1})^{1}(1-r^{2})^{2}(1-r^{3})^{3}\\\nonumber
Q_{\mathrm{per}\,\tilde{C}_{3}=1}(r)&=&\displaystyle \prod_{m=1}^{3}(1-r^{m})^{W_{3}(m)}=(1-r^{1})^{0}(1-r^{2})^{3}(1-r^{3})^{2}
\end{eqnarray}
The formulas for the exact probabilities $P_{\mathrm{per}\,\tilde{S}_{3}=u}(r)$ derived in~\cite{Ito2}   are
\begin{eqnarray}\nonumber
 P_{\mathrm{per}\,\tilde{A}_{3}=0}(r)&=&(1-r)^{9}+9r(1-r)^{8}+36r^{2}(1-r)^{7}+78r^{3}(1-r)^{6}\\\nonumber
&&+90r^{4}(1-r)^{5}+45r^{5}(1-r)^{4}+6r^{6}(1-r)^{3}\\\nonumber
 P_{\mathrm{per}\,\tilde{B}_{3}=0}(r)&=&(1-r)^{7}+6r(1-r)^{6}+13r^{2}(1-r)^{5}+10r^{3}(1-r)^{4}\\\nonumber
&&+2r^{4}(1-r)^{3}\\\nonumber
 P_{\mathrm{per}\,\tilde{C}_{3}=1}(r)&=&(1-r)^{6}+6r(1-r)^{5}+12r^{2}(1-r)^{4}+6r^{3}(1-r)^{3}\\\nonumber
\end{eqnarray}
All six of the functions given above (for $n=3$) are graphed vs.\ $r$ in Fig.~\ref{Fig6curves_n3_p1}.  
Also, Fig.~\ref{Fig6curves_n5_p1}  shows the six curves for $n=5$ vs.\ $r$.
For every $n\ge 1$ and all three types of matrices ($\tilde{\mathbf{A}}_{n},\\tilde{\mathbf{B}}_{n}$, and $\tilde{\mathbf{C}}_{n}$),
\begin{equation}
\label{EqUnVnWn_p1}
\displaystyle \sum_{m=0}^{n}E_{n}(m)=n!
\end{equation}
Thus the differences in the values of $E_{n}(m)$ for the different types of matrices stem entirely from differences in the distribution of $n!$ over the individual values of $m$. For type $\tilde{\mathbf{A}}_{n}$, that distribution is trivial (namely, $E_{n}(n)=n!$, and $E_{n}(m)=0$ for $m<n$). In the case of type $\tilde{\mathbf{C}}_n$, we found an analytic formula for the distribution:\[\displaystyle W_{n}(m)={}_{n}\mathrm{P}_{m} \cdot\sum_{l=0}^{m}\frac{(-1)^{l}}{l!},\qquad m=0,1,2,\ldots,n\]As illustrated earlier, the distribution for type $\tilde{\mathbf{C}}_{n}$ can also be derived from the cycle structure of the permutation group $\mathfrak{S}_n$. Moreover, we have shown that the distribution for type $\tilde{\mathbf{B}}_{n}$ can be derived from that for type $\tilde{\mathbf{C}}_{n}$, by using the relation $V_{n}(m)-W_{n-1}(m-1)=W_{n}(m)-W_{n-1}(m)$. 
As established in~\cite{Ito2}, there is an analytic formula for the coefficients in $P_{\mathrm{per}\,\tilde{C}_{n}=1}(r)$, since they can be derived from characteristics of acyclic digraphs with vertex set $\{1,2,3,\ldots,n\}$. For several values of $n$, the coefficients in $P_{\mathrm{per}\,\tilde{A}_{n}=0}(r)$ and $P_{\mathrm{per}\,\tilde{B}_{n}=0}(r)$ were also presented in that paper, but they were determined by computer, using an algorithm that took a binary matrix of a given type as input and computed its permanent. It remains to be seen whether there are explicit formulas for the coefficients in $P_{\mathrm{per}\,\tilde{A}_{n}=0}(r)$ and $P_{\mathrm{per}\,\tilde{B}_{n}=0}(r)$, and whether there is an analytic formula that relates one set of coefficients to the other.

\end{document}